\documentclass[conference]{IEEEtran}
\usepackage{amsmath}
\usepackage{latexsym}
\usepackage{graphicx}
\usepackage{bbding}
\usepackage{indentfirst}
\IEEEoverridecommandlockouts
\begin{document}

\title{Optimal Power Allocation for Secure Communications in Large-Scale MIMO Relaying Systems}
\author{\authorblockN{Jian~Chen$^{\dagger}$, Xiaoming~Chen$^{\dagger}$,
Xiumin~Wang$^{\ddagger}$, and Lei~Lei$^{\dagger}$
\\$^{\dagger}$College of Electronic and Information Engineering, Nanjing
University of Aeronautics and Astronautics, China.
\\$^{\ddagger}$School of Computer and Information, Hefei University of
Technology, Hefei, China.
\\Email: chenxiaoming@nuaa.edu.cn
%\thanks{This work was supported by the Natural
%Science Foundation of China (No. 61100195, 61301102), the Natural
%Science Foundation of Jiangsu Province (No. BK20130820), and the
%Doctoral Fund of Ministry of Education of China (No.
%20123218120022).}
}} \maketitle

\begin{abstract}
In this paper, we address the problem of optimal power allocation at
the relay in two-hop secure communications. In order to solve the
challenging issue of short-distance interception in secure
communications, the benefit of large-scale MIMO (LS-MIMO) relaying
techniques is exploited to improve the secrecy performance
significantly, even in the case without eavesdropper channel state
information (CSI). The focus of this paper is on the analysis and
design of optimal power allocation for the relay, so as to maximize
the secrecy outage capacity. We reveal the condition that the
secrecy outage capacity is positive, prove that there is one and
only one optimal power, and present an optimal power allocation
scheme. Moreover, the asymptotic characteristics of the secrecy
outage capacity is carried out to provide some clear insights for
secrecy performance optimization. Finally, simulation results
validate the effectiveness of the proposed scheme.
\end{abstract}

\section{Introduction}
Wireless security is always a critical issue due to the open nature
of the wireless channel. Traditionally, high-layer encryption
techniques are adopted to guarantee secure communications. However,
information-theoretic study shows that the originally harmful
factors of wireless channels, such as fading, noise and
interference, can be used to realize wireless security, namely
physical layer security \cite{Shannon} \cite{Wyner}, then the
complicated encryption can be partially replaced, especially in
mobile communications.

It has been proved repeatedly that the secrecy performance is
determined by the rate difference between the legitimate channel and
the eavesdropper channel \cite{SC1} \cite{SC2}. To improve the
secrecy performance, multi-antenna relaying techniques are commonly
used in physical layer security \cite{Relay}. On the one hand, the
use of the relay shortens the access distance, and thus increases
the legitimate channel rate. On the other hand, multi-antenna
techniques can be applied to impair the interception signal. The
beamforming schemes at the MIMO relay based on global channel state
information (CSI) for amplify-and-forward (AF) and
decode-and-forward (DF) relaying systems were presented in \cite{AF}
and \cite{DF}, respectively. Note that the beam design in secure
communications requires both legitimate and eavesdropper CSI
\cite{CSI}. However, it is usually difficult to obtain eavesdropper
CSI due to the well hidden property of the eavesdropper. In this
context, the beam is not optimal, and thus the secrecy performance
is degraded. To solve it, a joint jamming and beamforming scheme at
the relay in the case without eavesdropper CSI was proposed in
\cite{Jamming}. The relay transmits the artificial noise signal in
the null space of the legitimate channel together with the forward
signal, so the quality of the interception signal is weakened. This
scheme improves the secrecy performance at the cost of power
efficiency.

Recently, LS-MIMO relaying techniques are introduced into secure
communications to improve the secrecy performance
\cite{LS-MIMORelaying}. It is found that even without eavesdropper
CSI, LS-MIMO techniques can produce a high-resolution spatial beam,
then the information leakage to the eavesdropper is quite small.
More importantly, the secrecy performance can be enhanced by simply
adding the antennas. Thus, the challenging issue of short-distance
interception in secure communications can be well solved. Note that
in two-hop secure systems, the transmit power at the relay has a
great impact on the secrecy performance, since the power will affect
the signal quality at the destination and the eavesdropper
simultaneously. An optimal power allocation scheme for a
multi-carrier two-hop single-antenna relaying network was given by
maximizing the sum secrecy rate in \cite{PowerAllocation}. However,
the power allocation for a multi-antenna relay, especially an
LS-MIMO relay, is still an open issue. In this paper, we focus on
power allocation for secure two-hop LS-MIMO relaying systems under
very practical assumptions, i.e., no eavesdropper CSI and imperfect
legitimate CSI. The contributions of this paper are three-fold:

\begin{enumerate}
\item We reveal the relation between the secrecy outage capacity
and the defined relative distance-dependent path loss, and then give
the condition that the secrecy outage capacity is positive.

\item We prove that there is one and only one optimal power at the
relay, and propose an optimal power allocation scheme.

\item We present several clear insights for secrecy performance optimization
through asymptotic analysis.
\end{enumerate}

The rest of this paper is organized as follows. We first give an
overview of the secure LS-MIMO relaying system in Section II, and
then analyze and design an optimal power allocation scheme for the
relay in Section III. In Section IV, we present some simulation
results to validate the effectiveness of the proposed scheme.
Finally, we conclude the whole paper in Section V.

\section{System Model}
\begin{figure}[h] \centering
\includegraphics [width=0.4\textwidth] {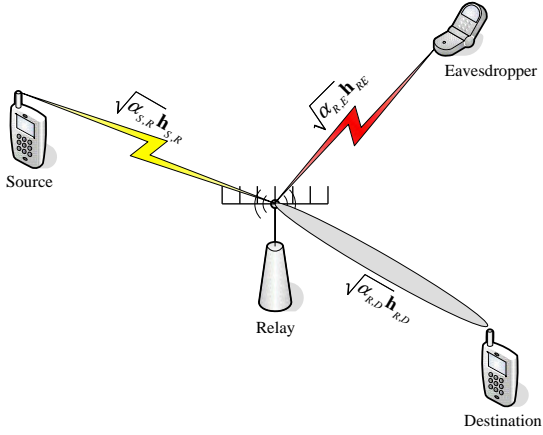}
\caption {An overview of a secure LS-MIMO relaying system.}
\label{Fig1}
\end{figure}

Consider a time division duplex (TDD) two-hop LS-MIMO relaying
system, as shown in Fig.1. It consists of one source, one
destination and one passive eavesdropper, equipped with a single
antenna each, and one relay with  $N_R$ antennas. It is worth
pointing out that $N_R$ is quite large in this LS-MIMO relaying
system, i.e. $N_R=100$ or larger. In addition, it is assumed that
the distance between the source and the destination is so long that
it is impossible to transmit the information from the source to the
destination directly. The whole system works in a half-duplex mode,
which means that a complete transmission requires two time slots.
Specifically, in the first time slot, the source sends the signal to
the relay, and then the relay forwards the post-processing signal to
the destination during the second time slot. We assume that the
eavesdropper is far away from the source and close to the relay,
since it thought the signal comes from the relay. Then, the
eavesdropper only monitors the transmission from the relay to the
destination. Note that this is a common assumption in previous
related literatures, because it is difficult for the eavesdropper to
monitor both the source and the relay.

We use $\sqrt{\alpha_{S,R}}\textbf{h}_{S,R}$,
$\sqrt{\alpha_{R,D}}\textbf{h}_{R,D}$ and $\sqrt{\alpha_
{R,E}}\textbf{h}_{R,E}$ to represent the channels from the source to
the relay, the relay to the destination, and the relay to the
eavesdropper respectively, where $\alpha_{S,R}$, $\alpha_{R,D}$ and
$\alpha_{R,E}$ are the distance-dependent path losses and
$\textbf{h}_{S,R}$, $\textbf{h}_{R,D}$, and $\textbf{h}_{R,E}$ are
channel small scale fading vectors with independent and identically
distributed (i.i.d.) zero mean and unit variance complex Gaussian
entries. It is assumed that the channels remain constant during a
time slot and fade independently over slots. Thus, the received
signal at the relay in the first time slot can be expressed as
\begin{equation}
\textbf{y}_R=\sqrt{P_S\alpha_{S,R}}\textbf{h}_{S,R}s+\textbf{n}_R,\label{eqn1}
\end{equation}
where $s$ is the normalized Gaussian distributed transmit signal,
$P_S$ is the transmit power at the source, $\textbf{n}_R$ is the
additive Gaussian white noise with zero mean and unit variance at the relay.

Then, the relay adopts an amplify-and-forward (AF) relaying protocol
to forward the received signal. Due to the low complexity and good
performance in LS-MIMO systems, we combine maximum ratio combination
(MRC) and maximum ratio transmission (MRT) at the relay to process
the received signal. We further assume that the relay has perfect
CSI about $\textbf{h}_{S,R}$ by channel estimation and gets partial
CSI about $\textbf{h}_{R,D}$ due to channel reciprocity in TDD
systems. The relation between the estimated CSI
$\hat{\textbf{h}}_{R,D}$ and the real CSI $\textbf{h}_{R,D}$ is
given by
\begin{equation}
\textbf{h}_{R,D}=\sqrt{\rho}\hat{\textbf{h}}_{R,D}+\sqrt{1-\rho}\textbf{e},\label{eqn2}
\end{equation}
where $\textbf{e}$ is the error noise vector with i.i.d. zero mean
and unit variance complex Gaussian entries, and is independent of
$\hat{\textbf{h}}_{R,D}$. $\rho$, scaling from $0$ to $1$, is the
correlation coefficient between $\hat{\textbf{h}}_{R,D}$ and
$\textbf{h}_{R,D}$. Then, the normalized signal to be transmitted at
the relay can be expressed as
\begin{equation}
\textbf{r}^{AF}=\textbf{F}\textbf{y}_{R},\label{eqn3}
\end{equation}
where \textbf{F} is the processing matrix, which is given by
\begin{equation}
\textbf{F}=\frac{\hat{\textbf{h}}_{R,D}}{\|\hat{\textbf{h}}_{R,D}\|}\frac{1}{\sqrt{P_S\alpha_{S,R}\|\textbf{h}_{S,R}\|^2+1}}\frac{\textbf{h}_{S,R}^H}{\|\textbf{h}_{S,R}\|}.\label{eqn4}
\end{equation}

Thus, the received signals at the destination and the eavesdropper
are given by
\begin{equation}
y_D=\sqrt{P_R\alpha_{R,D}}\textbf{h}_{R,D}^H\textbf{r}^{AF}+n_D,\label{eqn5}
\end{equation}
and
\begin{equation}
y_{E}=\sqrt{P_R\alpha_{R,E}}\textbf{h}_{R,E}^H\textbf{r}^{AF}+n_{E},\label{eqn6}
\end{equation}
respectively, where $P_R$ is the transmit power of the relay, $n_D$
and $n_{E}$ are the additive Gaussian white noises with zero mean
and unit variance at the destination and the eavesdropper.

Since there is no knowledge of the eavesdropper channel at the
source and the relay, it is impossible to provide a steady secrecy
rate over all realizations of the fading channels. In this paper, we
take the secrecy outage capacity $C_{SOC}$ as the performance
metric, which is defined as the maximum available rate under the
condition that the outage probability that the real transmission
rate surpasses the secrecy rate is equal to a given value
$\varepsilon$, namely
\begin{equation}
P_r(C_{SOC}>C_D-C_E)=\varepsilon,\label{eqn7}
\end{equation}
where $C_D$ and $C_E$ are the legitimate and the eavesdropper
channel rates, respectively.

Note that $C_{SOC}$ is not an decreasing function of $P_R$, since
both $C_D$ and $C_E$ increase as $P_R$ adds. Then, it makes sense to
select an optimal $P_R$. The focus of this paper is on the optimal
power allocation at the relay, so as to maximize the secrecy outage
capacity for a given outage probability.

\section{Optimal Power Allocation}
In this section, we first analyze the condition that the secrecy
outage capacity is positive, prove the existence of one and only one
optimal power, and then design an optimal power allocation scheme
for the relay. Finally, we present the asymptotic characteristics of
the secrecy outage capacity.

Note that accurate performance analysis is the basis of power
allocation. Prior to designing the optimal power allocation scheme,
we first reveal the relation between the secrecy outage capacity and
the transmit power. Based on the received signals in (\ref{eqn3})
and (\ref{eqn4}), the signal-to-noise ratio (SNR) at the destination
and the eavesdropper can be expressed as
\begin{equation}
\gamma_D=\frac{P_SP_R\alpha_{S,R}\alpha_{R,D}|\textbf{h}_{R,D}^H\hat{\textbf{h}}_{R,D}|^2\|\textbf{h}_{S,R}\|^2}{P_R\alpha_{R,D}|\textbf{h}_{R,D}^H\hat{\textbf{h}}_{R,D}|^2+\|\hat{\textbf{h}}_{R,D}\|^2(P_S\alpha_{S,R}\|\textbf{h}_{S,R}\|^2+1)},\label{eqn8}
\end{equation}
and
\begin{equation}
\gamma_E=\frac{P_SP_R\alpha_{S,R}\alpha_{R,E}
|\textbf{h}_{R,E}^H\hat{\textbf{h}}_{R,D}|^2\|\textbf{h}_{S,R}\|^2}
{P_R\alpha_{R,E}|\textbf{h}_{R,E}^H\hat{\textbf{h}}_{R,D}|^2
+\|\hat{\textbf{h}}_{R,D}\|^2(P_S\alpha_{S,R}\|\textbf{h}_{S,R}\|^2+1)}.\label{eqn9}
\end{equation}
Then, the legitimate and the eavesdropper channel rates are given by
$C_D=W\log_2(1+\gamma_D)$ and $C_E=W\log_2(1+\gamma_E)$
respectively, where $W$ is a half of the spectral bandwidth, since a
complete transmission requires two time slots. Thus, for the secrecy
outage capacity, we have the follow lemma:

\emph{Lemma 1}: For a given outage probability by $\varepsilon$, the
secrecy outage capacity of an LS-MIMO relaying system with imperfect
CSI can be expressed as
$C_{SOC}=W\log_2\left(1+\frac{P_SP_R\alpha_{S,R}\alpha_{R,D}\rho
N_R^2}{P_R\alpha_{R,D}\rho N_R+P_S\alpha_{S,R}N_R+1}\right)
-W\log_2\left(1+\frac{P_SP_R\alpha_{S,R}\alpha_{R,E}N_R\ln\varepsilon}{P_R\alpha_{R,E}\ln\varepsilon-P_S\alpha_{S,R}N_R}-1\right)$.

\begin{proof}
The secrecy outage capacity can be obtained based on (\ref{eqn7}) by
making use of the property of channel hardening in LS-MIMO systems
\cite{ChannelHardening}. We omit the proof, and the detail can be
referred to our previous work \cite{LS-MIMORelaying}.
\end{proof}

\subsection{Positiveness}
It is worth pointing out that the secrecy outage capacity may be
negative or zero from a pure mathematical view. Therefore, it makes
sense to find the condition that the positive secrecy outage
capacity exists.

Let $\rho\alpha_{R,D}N_R=A$, $-\alpha_{R,E}\ln\varepsilon=A\cdot
r_{gl}$, $P_S\alpha_{S,R}N_R=B$, where
$r_{l}=\frac{-\alpha_{R,E}\ln\varepsilon}{\rho\alpha_{R,D}N_R}$ is
defined as the relative distance-dependent path loss. Then, the
secrecy outage capacity can be rewritten as
\begin{eqnarray}
C_{SOC}&=&W\log_2\left(1+\frac{P_RAB}{P_RA+B+1}\right)\nonumber\\
&&-W\log_2\left(1+\frac{P_RABr_{l}}{P_RAr_{l}+B+1}\right).\label{eqn12}
\end{eqnarray}

Observing the secrecy outage capacity in (\ref{eqn12}), we get the
following theorem:

\emph{Theorem 1}: If and only if $0<r_{l}<1$, the secrecy outage
capacity in an LS-MIMO relaying system in presence of imperfect CSI
is positive.

\begin{proof}
Please refer to Appendix I.
\end{proof}

\emph{Remarks}: It is known that from Theorem 1, $0<r_l<1$ is a
precondition for power allocation in such an LS-MIMO relaying
system. Given channel conditions and outage probability, there is a
constraint on the minimum number of antennas at the relay in order
to fulfill $0<r_l<1$. Then, we have the following proposition:

\emph{Proposition 1}: The number of antennas $N_R$ at the relay must
be greater than
$\frac{-\alpha_{R,E}\ln\varepsilon}{\rho\alpha_{R,D}}$.

Note that even with a stringent requirement on the outage
probability, $\frac{-\alpha_{R,E}\ln\varepsilon}{\rho\alpha_{R,D}}$
can be always met by adding the antennas, which is an advantage of
an LS-MIMO relaying system. In what follows, we only consider the
case of $0<r_{l}<1$.

\subsection{Existence and Uniqueness}
As shown in (\ref{eqn12}), the secrecy outage capacity is not an
increasing function of $P_R$. Then, there may be an optimal power
for the relay in the sense of maximizing the secrecy outage
capacity. In this subsection, we aim to prove that the optimal power
exists and is unique.

Prior to seeking the optimal power, we first check two extreme cases
of $P_R$. On the one hand, if $P_R$ is large enough, the terms $B+1$
in (\ref{eqn12}) is negligible, so the secrecy outage capacity is
reduced as $C_{SOC}=W\log_2\left(1+\frac{P_RAB}{P_RA}\right)
-W\log_2\left(1+\frac{P_RABr_{l}}{P_RAr_{l}}\right)=0$. In other
words, when $P_R$ is very large, the SNRs at the destination and the
eavesdropper asymptotically approach the same value. Thus, the
secrecy outage capacity becomes zero. On the other hand, when $P_R$
tends to zero, the secrecy outage capacity is equal to
$C_{SOC}=W\log_2\left(1+\frac{0}{B+1}\right)-W\log_2\left(1+\frac{0}{B+1}\right)
=0$. Under this situation, both the rates of legitimate and
eavesdropper channels tend to zero, and thus the secrecy outage
capacity is also zero.

According to Theorem 1, the secrecy outage probability is positive
when $0<r_l<1$, so the maximum secrecy outage capacity must appear
at medium $P_R$ regime. Then, we get the following theorem:

\emph{Theorem 2}: From the perspective of maximizing the secrecy
outage capacity, the optimal power at the relay in an LS-MIMO
relaying system exists and is unique, once the relative
distance-dependent path loss $r_l$ is less than 1.

\begin{proof}
Please refer to Appendix II.
\end{proof}

\subsection{Optimal Power Allocation}
From Theorem 2, it is known that as long as $0<r_{l}<1$, there is
always a unique optimal power. In other words, if the relay applies
the optimal power, the LS-MIMO relaying system gets the maximum
secrecy outage capacity. Then, we have the following theorem:

\emph{Theorem 3}: When the relay uses the power
$P_R^{\star}=\sqrt{\frac{P_S\alpha_{S,R}N_R+1}{-\alpha_{R,E}\rho\alpha_{R,D}N_R\ln\varepsilon}}$,
the LS-MIMO relaying system gets the maximum secrecy outage
capacity, which is given by
$C_{SOC}^{\max}=W\log_2\left(1+\frac{P_S\alpha_{S,R}N_R}{1+\sqrt{\frac{-\alpha_{R,E}\ln\varepsilon}{\rho\alpha_{R,D}N_R}(1+P_S\alpha_{S,R}N_R)}}\right)-W\log_2\left(1+\frac{P_S\alpha_{S,R}N_R}{1+\sqrt{\frac{-\rho\alpha_{R,D}N_R}{\alpha_{R,E}\ln\varepsilon}(1+P_S\alpha_{S,R}N_R)}}\right)$.

\begin{proof}
Substituting the optimal power $P_R$ in (\ref{eqn20}) into $C_{SOC}$
in (\ref{eqn12}), we can derive the maximum secrecy outage capacity.
\end{proof}

\emph{Remarks}: The optimal power at the relay $P_R^{\star}$ is an
increasing function of source transmit power $P_S$, source-relay
path loss $\alpha_{S,R}$ and outage probability $\varepsilon$, and
is a decreasing function of CSI accuracy $\rho$, relay-destination
path loss $\alpha_{R,D}$ and relay-eavesdropper path loss
$\alpha_{R,D}$. In addition, due to
$r_{l}=\frac{-\alpha_{R,E}\ln\varepsilon}{\rho\alpha_{R,D}N_R}<1$,
the maximum secrecy outage capacity is an increasing function of
$P_S$, $\alpha_{S,R}$, $\alpha_{R,D}$, $\varepsilon$, $N_R$ and
$\rho$, and is a decreasing function of $\alpha_{R,E}$.

\subsection{Asymptotic Characteristic}
As analyzed above, the optimal power at the relay $P_R^{\star}$ is
an increasing function of the power at the source $P_S$. Next, we
carry out asymptotic analysis to $P_S$ and get the following
theorem:

\emph{Theorem 4}: At the low $P_S$ regime, the optimal power
$P_R^{\star}$ and the maximum secrecy outage capacity
$C_{SOC}^{\max}$ tend to zero. In the high $P_S$ region, the maximum
secrecy outage capacity will be saturated and is independent of
$P_S$.

\begin{proof}
Please refer to Appendix III.
\end{proof}

As $P_S$ approaches zero, the source does not transmit any
information to the relay in the first slot, so the maximum secrecy
outage capacity tends to zero. While $P_S$ is sufficiently large,
the forward noise at the relay is also amplified, and thus the
secrecy outage capacity is saturated and is independent of $P_S$ and
$P_R$.

\section{Simulation Results}
To examine the effectiveness of the proposed optimal power
allocation scheme for the AF LS-MIMO relaying system, we present
several simulation results in the following scenarios: we set
$N_R=100$, $W=10$KHz, $\rho=0.9$ and $\varepsilon=0.01$. We assume
that the relay is in the middle of the source and the destination.
For convenience, we normalize the pass loss as
$\alpha_{S,R}=\alpha_{R,D}=1$ and use $\alpha_{S,E}$ to denote the
relative path loss. Specifically, $\alpha_{R,E}>1$ means the
eavesdropper is closer to the relay than the destination. We use
SNR$_S=10\log_{10}P_S$ and SNR$_R=10\log_{10}P_R$ to represent the
transmit signal-to-noise ratio (SNR) in dB at the source and the
relay, respectively.

\begin{figure}[h] \centering
\includegraphics [width=0.5\textwidth] {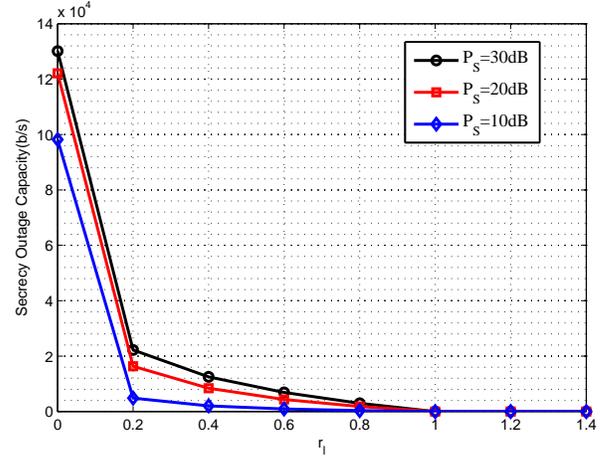}
\caption {Secrecy outage capacity with different relative
distance-dependent path losses.} \label{Fig2}
\end{figure}

First, we show the impact of $r_{l}$ on the secrecy outage capacity
with SNR$_R=20$dB. As seen in Fig.\ref{Fig2}, the positive secrecy
outage capacity exists only when $0<r_{l}<1$, which confirms the
claims in Theorem 1. Given a $r_l$, the secrecy outage capacity
increases gradually as $P_S$ adds. However, the performance loss by
reducing $P_S$ from 30dB to 20dB is smaller than that by reducing
$P_S$ from 20dB to 10dB. This is because in the large $P_S$ region,
the secrecy outage capacity tends to be saturated.

\begin{figure}[h] \centering
\includegraphics [width=0.5\textwidth] {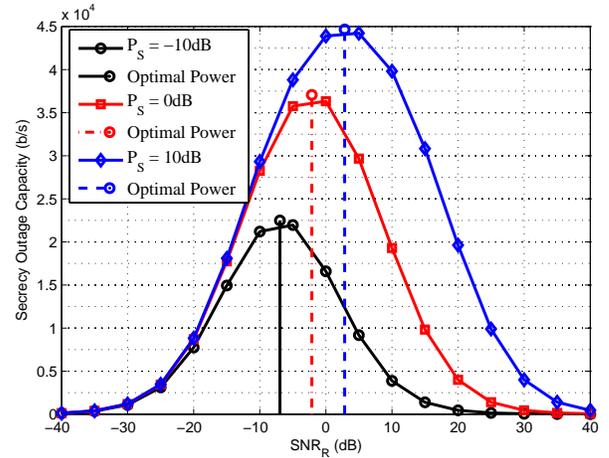}
\caption {Secrecy capacity with different SNR$_R$.} \label{Fig3}
\end{figure}

Second, we validate the existence and uniqueness of the optimal
power $P_R^{\star}$. As showed in Fig.\ref{Fig3}, the secrecy outage
capacity approaches zero both when $P_S$ tends to zero and infinity,
and the unique optimal power associated to the maximum secrecy
outage capacity appears in the medium region of $P_S$. Furthermore,
it is found that both $P_R^{\star}$ and $C_{SOC}^{\max}$ improves as
$P_S$ increases, which confirms our theoretical claims again.

\begin{figure}[h] \centering
\includegraphics [width=0.5\textwidth] {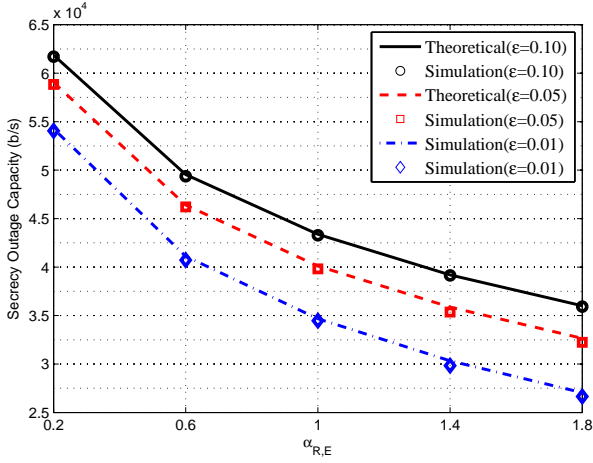}
\caption {Comparison of theoretical and simulation results.} \label{Fig4}
\end{figure}

Then, we testify the accuracy of the theoretical expression of the
maximum secrecy outage capacity with SNR$_S=10$dB. As seen in
Fig.\ref{Fig4}, the theorem results are well consistent with the
simulations in the whole $\alpha_{R,E}$ region with different outage
probability requirements, which proves the high accuracy of the
derived performance expression. As claimed above, given an outage
probability bound by $\varepsilon$, as $\alpha_{R,E}$ increases, the
maximum outage secrecy capacity decreases. This is because the
interception capability of the eavesdropper enhances when the
interception distance becomes small. What's more, given a
$\alpha_{R,E}$, the maximum secrecy outage capacity increases with
the increase of $\varepsilon$.

\begin{figure}[h] \centering
\includegraphics [width=0.5\textwidth] {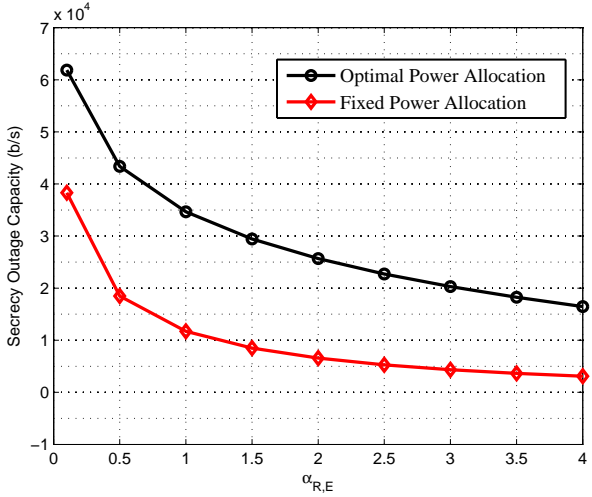}
\caption {Performance gain with different $\alpha_{R,E}$.} \label{Fig5}
\end{figure}

Next, we show the performance gain of the proposed optimal power
allocation scheme compared with a fixed power allocation scheme with
SNR$_S=10$dB. It is worth pointing out the fixed scheme uses a fixed
power $P_R=20$dB regardless of channel conditions and system
parameters. As seen in Fig.\ref{Fig5}, the optimal power allocation
scheme performs better than the fixed scheme. Even with a large
$\alpha_{R,E}$, such as $\alpha_{R,E}=4$, namely short-distance
interception, the optimal scheme can still achieve a high
performance gain, which proves the effectiveness of the proposed
scheme.

\begin{figure}[h] \centering
\includegraphics [width=0.5\textwidth] {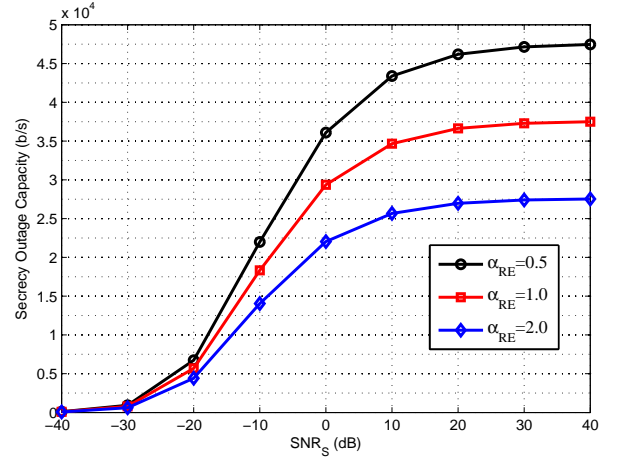}
\caption {Maximum secrecy capacity with different SNR$_S$.} \label{Fig6}
\end{figure}

Finally, we show the effect of $P_S$ on the maximum secrecy outage
capacity. As seen in Fig.\ref{Fig6}, when $P_S$ tends to zero, the
maximum secrecy outage capacity with different $\alpha_{R,E}$
approaches zero. In the large $P_S$ region, the maximum secrecy
outage capacity will be saturated for a given $\varepsilon$, which
proves the Theorem 3 again. Consistent with our theoretical
analysis, the performance ceiling is an decreasing function of
$\alpha_{R,E}$.

\section{Conclusion}
This paper focus on the optimal power allocation for a secure AF
LS-MIMO relaying system with imperfect CSI. We present the condition
that the secrecy outage capacity is positive, prove the existence
and uniqueness of the optimal power at the relay, and propose an
optimal power allocation scheme. Moreover, we reveal the asymptotic
characteristics of the maximum secrecy outage capacity in cases of
low and high source transmit powers.

\begin{appendices}
\section{Proof of Theorem 1}
To get the condition that the secrecy outage capacity is positive,
we first rewrite (\ref{eqn12}) as
\begin{eqnarray}
C_{SOC}&=&W\log_2\left(1+\frac{P_RAB}{P_RA+B+1}\right)\nonumber\\
&&-W\log_2\left(1+\frac{P_RAB}{P_RA+\frac{B+1}{r_{l}}}\right).\label{eqn17}
\end{eqnarray}
Examining (\ref{eqn17}), it is found that if and only if
$0<r_{l}<1$, the secrecy outage capacity is positive. According to
the definition of the relative distance-dependent path loss
$r_{l}=\frac{-\alpha_{R,E}\ln\varepsilon}{\rho\alpha_{R,D}N_R}$,
$0<r_{l}<1$ is equivalent to the following condition:
\begin{eqnarray}
N_R>\frac{-\alpha_{R,E}\ln\varepsilon}{\rho\alpha_{R,D}}.\label{eqn18}
\end{eqnarray}
In other words, only when
$N_R>\frac{-\alpha_{R,E}\ln\varepsilon}{\rho\alpha_{R,D}}$, the
secrecy outage capacity is positive. Therefore, we get Theorem 1 and
Proposition 1.

\section{Proof of Theorem 2}
At first, we take derivative of (\ref{eqn12}) with respect to $P_R$,
which is given by (\ref{equ19}) at the top of the next page.
\begin{figure*}
\begin{equation}
C_{soc}^{'}=\frac{W}{\ln2}B(1+B)\left(\frac{A}{(P_RA+B+1)^2+P_RAB(P_RA+B+1)}-\frac{Ar_{l}}{(P_RAr_{l}+B+1)^2+P_RABr_{l}(P_RAr_{l}+B+1)}\right).\label{equ19}
\end{equation}
\end{figure*}
Let $C_{soc}'=0$, we get two solutions
\begin{eqnarray}
P_R&=&\frac{1}{Ar_{l}}\sqrt{r_{l}(B+1)},\label{eqn20}
\end{eqnarray}
and
\begin{eqnarray}
P_R&=&-\frac{1}{Ar_{l}}\sqrt{r_{l}(B+1)}.\label{eqn21}
\end{eqnarray}

Considering $P_R>0$, (\ref{eqn20}) is the unique optimal solution in
this case. What's more, when
$P_R<\frac{1}{Ar_{l}}\sqrt{r_{l}(B+1)}$, we have $C_{soc}'>0$.
Otherwise, if $P_R>\frac{1}{Ar_{l}}\sqrt{r_{l}(B+1)}$, we have
$C_{soc}'<0$. Specifically, $C_{SOC}$ improves as $P_R$ increases in
the region from $0$ to $\frac{1}{Ar_{l}}\sqrt{r_{l}(B+1)}$, while
$C_{SOC}$ decreases as $P_R$ increases in the region from
$\frac{1}{Ar_{l}}\sqrt{r_{l}(B+1)}$ to infinity. Only when
$P_R=\frac{1}{Ar_{l}}\sqrt{r_{l}(B+1)}$, the secrecy outage capacity
achieves the maximum value. In other words, the optimal solution
exists and is unique. Hence, we get the Theorem 2.

\section{Proof of Theorem 4}
According to Theorem 3, the maximum secrecy outage capacity can be
expressed as
\begin{eqnarray}
C_{SOC}^{\max}&=&W\log_2\left(1+\frac{\sqrt{r_{l}(B+1)}B}{\sqrt{r_{l}(B+1)}+r_{l}(B+1)}\right)\nonumber\\
&&-W\log_2\left(1+\frac{\sqrt{r_{l}(B+1)}B}{\sqrt{r_{l}(B+1)}+(B+1)}\right),\nonumber\\
&=&W\log_2\left(1+\frac{1}{\frac{1}{B}+\sqrt{r_{l}(\frac{1}{B}+\frac{1}{B^2})}}\right)\nonumber\\
&&-W\log_2\left(1+\frac{1}{\frac{1}{B}+\sqrt{\frac{1}{r_{l}}(\frac{1}{B}+\frac{1}{B^2})}}\right).\label{eqn22}
\end{eqnarray}
Intuitively, $B$ tends to zero as $P_S$ approaches zero. Then,
$\frac{1}{\frac{1}{B}+\sqrt{r_{l}(\frac{1}{B}+\frac{1}{B^2})}}$ and
$\frac{1}{\frac{1}{B}+\sqrt{\frac{1}{r_{l}}(\frac{1}{B}+\frac{1}{B^2})}}$
in (\ref{eqn22}) becomes zero. Thus, we have $C_{SOC}^{\max}=0$. On
the other hand, if $P_S$ is large enough, $B$ is also very large.
Therefore, the maximum secrecy outage capacity is transformed as
\begin{eqnarray}
C_{SOC}^{\max}&=&W\log_2\left(1+\frac{\sqrt{r_{l}(B+1)}B}{\sqrt{r_{l}(B+1)}+r_{l}(B+1)}\right)\nonumber\\
&&-W\log_2\left(1+\frac{\sqrt{r_{l}(B+1)}B}{\sqrt{r_{l}(B+1)}+(B+1)}\right),\nonumber\\
&=&W\log_2\left(1+\frac{B}{1+\sqrt{r_{l}(B+1)}}\right)\nonumber\\
&&-W\log_2\left(1+\frac{B}{1+\sqrt{\frac{B+1}{r_{l}}}}\right),\nonumber
\end{eqnarray}
\begin{eqnarray}
&\approx&W\log_2\left(1+\frac{B}{\sqrt{r_{l}(B+1)}}\right)\nonumber\\
&&-W\log_2\left(1+\frac{B}{\sqrt{\frac{B+1}{r_{l}}}}\right)\label{eqn23}\\
&\approx&W\log_2\left(1+\frac{B}{\sqrt{r_{l}B}}\right)-W\log_2\left(1+\frac{B}{\sqrt{\frac{B}{r_{l}}}}\right)\label{eqn24}\\
&=&W\log_2\left(1+\sqrt{\frac{B}{r_{l}}}\right)-W\log_2\left(1+\sqrt{r_{l}B}\right),\nonumber\\
&=&W\log_2\left(\frac{\sqrt{\frac{B}{r_{l}}}}{\sqrt{r_{l}B}}\right),\nonumber\\
&=&W\log_2\left(\frac{1}{r_{l}}\right).\label{eqn25}\\
&=&W\log_2\left(\frac{\rho\alpha_{R,D}N_R}{-\alpha_{R,E}\ln\varepsilon}\right),\label{eqn26}
\end{eqnarray}
where (\ref{eqn23}) and (\ref{eqn24}) hold true because when $B$ is
big enough, the constant term $``1"$ is negligible. Hence, we get
the Theorem 3.
\end{appendices}

\end{document}